\documentclass[aps,prb,twocolumn,superscriptaddress,showpacs,amsmath,amssymb]{revtex4}

\usepackage{graphicx}
\usepackage[floatfix]{epsfig}
\usepackage{bm}

\begin{document}


\title{Precursor effects of the superconducting state
caused by $d$-wave phase-fluctuations above $T_c$}


\author{Thomas Eckl}
\email{eckl@physik.uni-wuerzburg.de}
\affiliation{Institut f\"ur Theoretische Physik und Astrophysik,
Universit\"at W\"urzburg, Am Hubland, D-97074 W\"urzburg, Germany}

\author{Werner Hanke}
\email{hanke@physik.uni-wuerzburg.de}
\affiliation{Institut f\"ur Theoretische Physik und Astrophysik,
Universit\"at W\"urzburg, Am Hubland, D-97074 W\"urzburg, Germany}

\date{\today}

\begin{abstract}

One of the hallmarks of high-temperature superconductors is a 
pseudogap regime appearing in the underdoped 
cuprates above the superconducting transition temperature $T_c$.
The pseudogap continously develops out of the superconducting gap.
In addition, high-frequency conductivity experiments 
show a superconducting scaling of the optical response
in the pseudogap regime, pointing towards a superconducting
origin of the pseudogap. The phase-fluctuation vortex
scenario is further supported by the measurement of an 
unusually large Nernst signal above $T_c$ and 
the recently observed field-enhanced diamagnetism which
scales with the Nernst signal. In this paper, we use a 
simple phenomenological model to calculate 
the paraconductivity and magnetic response
caused by phase fluctuations of the superconducting
order parameter
above $T_c$. Our results are in agreement with
experiments such as the superconducting scaling of the optical
response and the spin (or Pauli) susceptibility, and further
strengthen the idea of
a phase-fluctuation origin of the pseudogap.

\end{abstract}

\pacs{71.10.Fd, 71.27.+a, 74.72.-h}

\maketitle

\section{Introduction}
\label{intro}

A variety of different experiments indicate a suppression of
low-frequency spectral weight in the underdoped cuprates below a 
characteristic temperature $T^*$, that is higher than the 
superconducting (SC) transition temperature $T_c$ 
\cite{lo.sh.96,di.yo.96,di.no.97,ta.re.91,al.ma.93,or.th.90,re.re.98,mi.za.99,ku.fi.01,wa.on.02}.
The size of this pseudogap temperature regime scales with the
superconducting gap.
In addition, high-frequency conductivity
experiments in underdoped Bi$_2$Sr$_2$CaCu$_2$O$_{8+\delta}$ (Bi2212) 
\cite{co.ma.99} have indicated a 
SC scaling behavior of the optical conductivity.
Furthermore, a strongly enhanced Nernst signal was measured above
$T_c$ in underdoped samples of La$_{2-x}$Sr$_x$CuO$_4$ (LSCO), which is usually 
associated with the presence of vortices in the superconducting state 
and, therefore, implies that $T_c$ corresponds to a loss of phase rigidity 
rather than a vanishing of the pairing amplitude \cite{wa.on.02}.
Recently, a field-enhanced diamagnetism was found in the pseudogap state 
of Bi2212, which scales exactly like the Nernst signal
\cite{wa.li.05u} and further supports the vortex description for the loss of
phase coherence at $T_{c}$.

All these experiments point towards a picture, where the pseudogap
arises from phase fluctuations of the superconducting gap
\cite{em.ki.95}. In previous work \cite{ec.sc.02}, we provided a
detailed numerical
solution of a minimal model which, however, contains key ideas of the
phase-fluctuation scenario: In this scenario, below a ``mean-field''
temperature $T_{MF} \equiv T^{*}$, a $d_{x^{2}-y^{2}}$-wave gap
amplitude is assumed to develop. However, the SC transition is
suppressed to a considerably lower temperature $T_{c}$ by phase
fluctuations \cite{em.ki.95}. In the intermediate temperature regime
between $T^{*}$ and $T_{c}$, the phase fluctuations of the gap give rise
to the pseudogap phenomena. In this previous work, it was shown that a
two-dimensional 
BCS-like Hamiltonian with a $d_{x^2-y^2}$-wave gap and phase
fluctuations, which were treated by a Monte-Carlo (MC) simulation of an
$XY$ model, yields results
which compare very well with scanning tunneling measurements 
over a wide temperature range \cite{ec.sc.02,ku.fi.01}. Furthermore,
this phenomenological phase fluctuation model
was also able to explain the ``violation'' of the
in-plane optical integral in underdoped
Bi$_2$Sr$_2$CaCu$_2$O$_{8+\delta}$ (Bi2212) \cite{ec.ha.03} and the
characteristic change of the quasiparticle dispersion observed in
crossing $T_c$ in underdoped Bi2212 samples \cite{ec.ha.04}.

Here, we consider in detail two other experiments, which critically
probe the phase-fluctuation scenario. The first one is the optical
response. Our calculations were, in particular, motivated by optical
studies of the Berkeley group \cite{co.ma.99}. In this seminal work
high-frequency optical response, i.~e.~the complex
frequency dependent conductivity, was used to capture the short-time scale
dynamics of the phase correlations. Indeed, the pseudogap conjecture
had early on led to an intensive search for normal state ``remnants''
of the SC state, such as the infinite d.~c.~conductivity. However,
it was not before this Corson et. al. work that it was realized that
the short phase-correlation times require a high-frequency optical
probe to make the conductivity enhancement over normal-state values
observable. 

We give a detailed analysis of the temperature and frequency 
dependence of the optical conductivity
$\sigma(\omega)=\sigma_1(\omega)+i\sigma_2(\omega)$
within our phenomenological phase fluctuation model, which, for
completeness, is reintroduced in section II. In section \ref{paracon},
we compare the calculated paraconductivity due to $d$-wave phase fluctuations
above $T_c$ with the high-frequency conductivity experiments of
Ref.~\onlinecite{co.ma.99}. Salient features of these experiments are reproduced in
$\sigma(\omega)$ above $T_{c}$: at sufficiently high frequencies, the
phase-fluctuating state becomes indistinguishable from the SC state.
In this latter case $\sigma_{2} (\omega)$, i.~e.~the imaginary part,
shows the usual BCS-like behavior falling off as $1/\omega$, with a
prefactor being proportional to the phase-stiffness energy. This
generates in our calculations a scaling behavior of $\sigma_{2}
(\omega, T)$, in agreement with experiment, in that all $\omega\, \sigma_{2}
(\omega, T)$ curves collapse onto a single curve at the
Kosterlitz-Thouless (KT)-temperature $T_{c} = T_{KT}$. 

Since, in the phase fluctuation scenario, the pseudogap is due to
phase fluctuations of the SC order parameter, one might ask, how much
of the characteristic magnetic properties of a superconductor
are still observable in the pseudogap state of the underdoped cuprates.

These magnetic precursor effects of the ideal superconducting state
are twofold. Firstly, one would expect a form of fluctuating
diamagnetism which partially screens an applied magnetic field already
above $T_c$. Secondly, one would also expect that the paramagnetic
spin (or Pauli) susceptibility is reduced in a characteristic way 
below $T^*$ due to the formation of incoherent singlet-pairs, 
accompanied by the reduction of spectral weight at the Fermi surface. 

In experiment, the fluctuation-induced diamagnetism is only important
close to $T_c$. Since, at higher temperatures, the paramagnetic spin
response is typically one-order-of-magnitude larger than the
diamagnetic response, we concentrate here on the spin susceptibility
$\chi_s$. This latter quantity dominates the
T-dependence of the paramagnetic susceptibility in the experiments.

In section \ref{mag}, it is shown that the Pauli
spin susceptibility extracted from our phase-fluctuation model can
qualitatively account for the observed paramagnetic properties
over a wide range of temperatures in the pseudogap regime. Section V,
finally, presents a summary of our results.

\section{Phase-Fluctuation Model}

We consider the Hamiltonian \cite{ec.sc.02}
\begin{equation}
H=H_0+H_1,
\label{one}
\end{equation}
where $H_0$ is the usual tight-binding operator of
non-interacting electrons on a two-dimensional (2D) square lattice, i.~e.~
\begin{equation}
H_0=-t \sum_{\langle
  \vec{i}\,\vec{j}\rangle,\sigma}(c^\dagger_{\vec{i}\,\sigma}c_{\vec{j}\,\sigma}+c^\dagger_{\vec{j}\,\sigma}c_{\vec{i}\,\sigma})
- \mu \sum_{\vec{i},\sigma} n_{\vec{i}\,\sigma}.
\label{ham_0}
\end{equation}
Here, $c^\dagger_{\vec{i}\,\sigma}$ ($c_{\vec{i}\,\sigma}$) creates
(annihilates) an electron of spin $\sigma$ on the $\vec{i}^{\, th}$
site of the (2D) square lattice and  $n_{\vec{i}\,\sigma}=c^\dagger_{\vec{i}\,\sigma}c_{\vec{i}\,\sigma}$ is the number operator. $t$  denotes an effective
nearest-neighbor hopping-term and $\mu$ is the chemical potential. The
$\langle ij \rangle$ sum is over nearest-neighbor sites of the 2D
square lattice. For
simplicity of our minimal model, we set longer-ranged hoppings $t'$
equal to zero. All calculations were done for 10 \% doping, which
corresponds to a typical underdoped cuprate situation.

The second part of the Hamiltonian $H_1$ contains a BCS-like $d$-wave
interaction, which is given by
\begin{equation}
H_1=-g\sum_{\vec{i}\,\vec{\delta}}(\Delta_{\vec{i}\,\vec{\delta}}\langle\Delta_{\vec{i}\,\vec{\delta}}^\dagger\rangle 
+\Delta_{\vec{i}\,\vec{\delta}}^\dagger\langle\Delta_{\vec{i}\,\vec{\delta}}\rangle),
\label{ham_i_d}
\end{equation}
with $\vec{\delta}$ connecting site $i$ to its nearest-neighbor sites. 
The coupling constant $g$ stands for the strength of the effective
next-neighbor $d_{x^2-y^2}$-wave pairing-interaction.
The origin of this pairing interaction is not essential for the
further calculation. It can be either of purely electronic origin,
like spin fluctuations, or phonon mediated. The only
important point is, that there exists an effective pairing
interaction, that produces a finite $d_{x^2-y^2}$-wave gap amplitude as
one goes below a certain temperature $T^*$.
In contrast to conventional BCS theory, we consider the pairing-field amplitude
not as a constant real number, but rather as a complex quantity
\begin{equation}
\langle\Delta_{\vec{i}\,\vec{\delta}}^\dagger\rangle=
\frac{1}{\sqrt{2}}\langle c_{\vec{i}\,\uparrow}^\dagger c_{\vec{i}+\vec{\delta}\,\downarrow}^\dagger
-c_{\vec{i}\,\downarrow}^\dagger c_{\vec{i}+\vec{\delta}\,\uparrow}^\dagger\rangle=
\Delta\,e^{i \,\Phi_{\vec{i} \vec{\delta}}},
\label{pair1}
\end{equation}
with a {\sl constant} magnitude $\Delta$ and a {\sl fluctuating} bond-phase field $\Phi_{\vec{i} \vec{\delta}}$.
In order to get a description, where the {\sl center of mass}
phases of the Cooper pairs are the only relevant degrees of freedom
\cite{pa.ra.00}, the $d_{x^2-y^2}$-wave bond-phase field is written in
the following way
\begin{equation}
\Phi_{\vec{i} \vec{\delta}}=\left\{\begin{array}{l@{\quad \mathrm{for} \quad}l}
(\varphi_{\vec{i}} + \varphi_{\vec{i}+\vec{\delta}})/2 & \text{$\vec{\delta}$ in $x$-direction} \\
(\varphi_{\vec{i}} + \varphi_{\vec{i}+\vec{\delta}})/2 +\pi & \text{$\vec{\delta}$ in $y$-direction,} 
\end{array} \right.
\label{pair2}
\end{equation}
where $\varphi_{\vec{i}}$ is the {\sl center of mass} phase of a Cooper
pair localized at lattice site $\vec{i}$. 

To account for the proximity to the Mott insulating state
and thus the low superfluid density, we perform a {\sl quenched average}
over all possible phase configurations with the statistical weight
given by the classical $XY$ free energy
\begin{equation}
F\left[\varphi_i\right] = - J \sum_{\langle ij\rangle}
\cos\left(\varphi_i-\varphi_j\right).
\label{two}
\end{equation}
Here, the phase stiffness $J$ determines the
Kosterlitz-Thouless
transition temperature $T_{KT}$ to a quasi phase-ordered
state, a temperature, which we take as $T_c$ \cite{ec.sc.02,ec.ha.03,ec.ha.04}. 
Finally, we set $ T_c \approx \frac{1}{4} T^*$, where we have the STM
experiments of Ref.~\onlinecite{ku.fi.01} in mind, which corresponds to
a large pseudogap regime.

In Ref.~\onlinecite{ec.sc.02}, the justification for using a minimal model and,
in particular, the use of a classical $XY$ interaction in Eq.~(\ref{two}), has
been carefully discussed. Our belief, that this model contains key ideas of
the cuprate phase-fluctuation scenario, has been substantiated there
by a detailed numerical solution and the fact that characteristic and
crucial features of STM studies \cite{mi.za.99} could be reproduced
over a wide temperature range.

Recent work on a BCS-Hamiltonian with classical phase fluctuations can
also be found in Ref.~\onlinecite{ma.al.00}.

\section{Paraconductivity}

\label{paracon}

In a conventional BCS superconductor, superconducting (SC) fluctuations
with short-range phase coherence typically survive no more than $1$K
above $T_c$. 
Within a phase fluctuation scenario for the underdoped cuprates, one
would expect that pairing remains over a wide temperature range above
$T_c$, together with phase correlations which are of finite range in
space and time \cite{em.ki.95}.
Hence, although the system is in the normal state, signatures of the
ideal SC state should still be observable considerably above $T_c$,
{\sl if}  
the experiments probe short enough time scales. A likely candidate to
observe these effects are high-frequency conductivity experiments.

Indeed, microwave conductivity experiments on underdoped Bi2212
\cite{co.ma.99} were able to track the phase-correlation 
time in the normal state up to $25$K above $T_c$.
These experiments show, that the SC transition is ``smeared out''
over a considerable temperature range, when viewed at high-enough
frequencies (short-enough time scales). The imaginary part of the
conductivity finally disappears more than $25$K above $T_c \simeq 74$K
and shows a superconducting scaling behavior already above $T_c$.
The real part of the conductivity displays a characteristic
peak near $T_c$ at finite frequencies, on top of a ``background conductivity''
of normal
conducting electrons. This peak was interpreted by the authors 
of Ref.~\onlinecite{co.ma.99} as signature of the partially phase coherent electrons. 

All experiments were carried out with frequencies of a few hundred
GHz, which corresponds to $\omega \lesssim 0.01t$.
In order to compare these experiments with our phase fluctuation model,
some compromise has to be made, since 
finite-size effects become important below $\omega_{low} \simeq 0.03t$.
Therefore, we have chosen a set of frequencies
$\omega=\{0.1t,0.2t,0.3t,0.4t\}$, which are, on the one hand, much
larger than $\omega_{low}$, and on the other
hand much smaller than the relevant energy
scale, i.~e.~the SC gap size in our model
($\Delta_{sc}=1.0t$). Thus, in the following, one always should keep in
mind that the frequency closest to experiment is $\omega=0.1t$.

The optical conductivity from our phase-fluctuation model, i.~e.~
\cite{sc.wh.93}
\begin{equation}
\sigma_{xx}(\vec{k},\omega)= e^2 \;\frac{\langle k_x \rangle -
  \mathcal{D}_c(\vec{k},\omega+i\eta)}{i(\omega+i\eta)},
\label{sig_lat}
\end{equation}
was obtained by averaging the Matsubara current-current correlation
function
\begin{equation}
 \mathcal{D}_c(\vec{l}\,\tau,\,\vec{l}^{\prime}\,\tau^{\prime})=
-\text{Tr}\{ \hat{\rho}_G \, \text{T}_{\tau} [ j_x^p(\vec{l},\tau) \;
j_x^p(\vec{l}^{\prime},\tau^{\prime}) ]  \},
\label{mat_cc}
\end{equation}
and the operator for the 
local kinetic energy in $x$-direction
\begin{equation}
k_x(\vec{l})= -t \sum_{\sigma}
(c^\dagger_{\vec{l}+\vec{e}_x,\,\sigma} \, c_{\vec{l},\,\sigma}
+c^\dagger_{\vec{l},\,\sigma}\, c_{\vec{l}+\vec{e}_x,\,\sigma})
\label{kx}
\end{equation}
over all possible phase configurations using Eq.~(\ref{two}).
Here, $\hat{\rho}_{G}$ is the usual statistical operator, and
\begin{equation}
j_x^p(\vec{l})= it \sum_{\sigma}
(c^\dagger_{\vec{l}+\vec{e}_x,\,\sigma} \, c_{\vec{l},\,\sigma}
-c^\dagger_{\vec{l},\,\sigma}\, c_{\vec{l}+\vec{e}_x,\,\sigma})
\label{j_p_disc}
\end{equation}
denotes the paramagnetic current density operator in $x$-direction. 

\begin{figure}[t]
\begin{center}
\includegraphics[width=7cm]{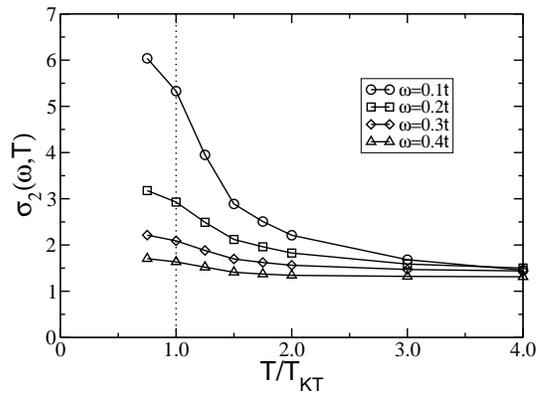}
\end{center}
\caption[]{Imaginary part of the low-frequency optical conductivity
  $\sigma_2(\omega,T)$, as a function of temperature for different
  frequencies $\omega < \Delta_{sc}$ ($\Delta_{sc}=1.0t$).
The dotted vertical line indicates $T_c \equiv T_{KT}$.
Note, that the change of  $\sigma_2(\omega,T)$ at the superconducting
transition $T_c \equiv T_{KT}$
becomes less pronounced at higher frequencies and is smeared out over
a finite range of temperatures $T>T_c\equiv T_{KT}$.}
\label{s2_temp}
\end{figure}

\begin{figure}[t]
\begin{center}
\includegraphics[width=7cm]{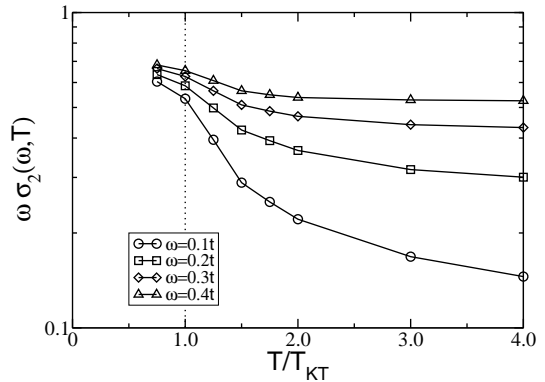}
\end{center}
\caption[]{Scaling behavior of the imaginary part of the optical
  conductivity  $\sigma_2(\omega,T)$. In the superconducting state
$\sigma_2(\omega,T) \sim 1/\omega$, thus all $\omega \;
\sigma_2(\omega,T)$ curves should collapse onto a single curve.
Note, that the higher the frequency, the earlier this collapse starts  
in the normal state for temperatures $T>T_c\equiv T_{KT}$.
The logarithmic scale was chosen for a better comparison with the
 experimental data of Ref.~\onlinecite{co.ma.99}, reproduced in Fig. 3,
 below.}
\label{s2_om_temp}
\end{figure}

\begin{figure}[t]
\begin{center}
\includegraphics[width=7cm]{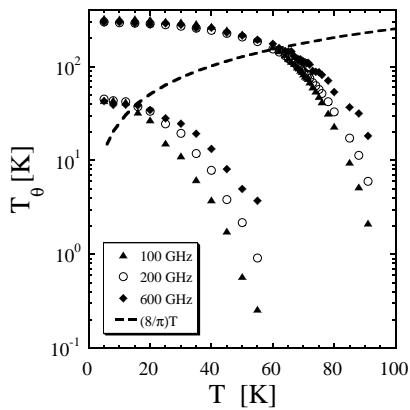}
\end{center}
\caption[]{Experimental results for the dynamic (frequency dependent)
phase-stiffness temperature $T_{\theta} \sim \omega \;
\sigma_2(\omega,T)$ in the pseudogap state of underdoped 
Bi2212 from Ref.~\onlinecite{co.ma.99}. Data are shown for two samples,
one with $T_c=33K$ (left side) and the other with $T_c=74K$ (right side).
The three curves for each sample correspond to measurement frequencies of
100, 200 and 600 GHz. (For details, see text.)}
\label{phase_fig}
\end{figure}

Fig.~\ref{s2_temp} plots our results for the imaginary part of the 
optical conductivity,i.e. $\sigma_2(\omega,T)$, as a function of 
temperature for different frequencies $\omega < \Delta_{sc}$.
Exactly as in the experiments of Ref.~\onlinecite{co.ma.99},
the superconductivity induced change in the imaginary part extends over a large temperature 
interval $T_c \lesssim T \lesssim 2 \, T_c$ into the normal state,
with the strongest increase below  $T \simeq 1.5 \, T_c$. 
The change of $\sigma_2(\omega,T)$ is more pronounced at 
lower frequencies. However, in our model the imaginary part does not
go to zero in the normal state but rather converges towards a finite
value. This is due to the missing quasiparticle scattering in
the normal state in addition to phase-fluctuation caused scattering.
In a ``real'' metal, above $T_c$, the finite electronic scattering
rate, strongly suppresses the imaginary part $\sigma_2 (\omega)$
at low frequencies, so that it can be neglected. This effect can
phenomenologically be taken into account by replacing the
infinitesimal damping factor $\eta$ in Eq.~(\ref{sig_lat}) by a finite
marginal-Fermi-liquid (MFL) damping factor \cite{li.va.92,eckl.04}. Since the
inclusion of such a MFL scattering-rate further reduces
$\sigma_2(\omega,T)$ above $T \simeq 2 \, T_c$, \emph{however without
	changing its properties at the SC transition ($T \lesssim 2 \, T_c$)
qualitatively,}
we refrain from burdening our simple model with it.

In Fig.~\ref{s2_om_temp}, the re-scaled imaginary part of the
optical conductivity $\omega \, \sigma_2(\omega,T)$ is displayed as a
function of temperature for the same frequencies $\omega <
\Delta_{sc}$, as before.
In the superconducting state, all  re-scaled curves 
should collapse onto a single curve, as has been beautifully
demonstrated in Ref.~\onlinecite{co.ma.99} (Fig.~2 of that work, which corresponds to
our Fig.~\ref{s2_om_temp} is reproduced for comparison in Fig.~\ref{phase_fig}).
One can clearly see from our results that this collapse already begins in the normal
state above $T_c \equiv T_{KT}$, starting with the highest
frequencies, exactly as in Ref.~\onlinecite{co.ma.99}. Also here, the
inclusion of an additional MFL scattering-rate reduces the optical conductivity curves above $T \simeq 2 \, T_c$,
but it does not change the SC high-frequency scaling-behavior 
below $T \lesssim 1.5 \, T_c$ \emph{significantly} \cite{eckl.04}.

\section{Magnetic properties}

\label{mag}

In this section, we study the precursor effects of the SC state
caused by d-wave phase fluctuations above $T_{c}$ appearing in the paramagnetic response.
The spin (or Pauli) susceptibility can be obtained from the coupling
of the magnetic field $\vec{B}$ to the spin degrees of freedom. With
$\vec{B} \parallel \hat{\vec{z}}$,
one gets \cite{whit.83}
\begin{equation}
\chi_{spin}(\vec{q},\omega)=
-\frac{N}{V} \; g^2\; \frac{\mu_B^2}{\hbar^2}\;\frac{1}{\hbar}\;
D_{zz}(\vec{q},\omega),
\end{equation}
where $D_{zz}(\vec{q},\omega)$ is the Fourier transform of the spin-spin
correlation function in $z$-direction, i.~e.~
\begin{equation}
D_{zz}(\vec{x},t,\,\vec{x}^\prime,t^\prime)=-i\; \text{Tr} \{
\hat{\rho}_G [s_z(\vec{x},t),s_z(\vec{x}^{\prime},t^{\prime})]\}
\,\Theta(t-t^{\prime}).
\end{equation}
The static uniform susceptibility is defined by
$\chi_{spin}=\chi_{spin}(\vec{q}\rightarrow 0,\omega=0)$.
For a non-interacting system with a general dispersion
$\epsilon(\vec{k})$, the static uniform  spin (or Pauli) susceptibility
is given by
\begin{equation}
\chi_{spin}(\vec{q}\rightarrow 0,\omega=0)=
-\frac{1}{V}\;\frac{1}{2} \; g^2\; \mu_B^2 \;
\sum_{\vec{k}}\, \frac{\partial f(\epsilon(\vec{k})-\mu)}{\partial \epsilon(\vec{k})},
\end{equation}
with  $f(\epsilon(\vec{k})-\mu)$ being the Fermi function.
For $T \rightarrow 0$, the static uniform susceptibility is proportional to the density of states at
the Fermi surface.

\begin{figure}[t]
\begin{center}
\includegraphics[width=7cm]{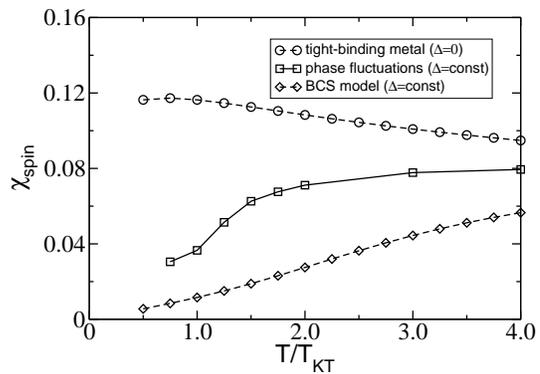}
\end{center}
\caption[]{Spin susceptibility $\chi_{spin}$ in units of $g^2 \mu_B^2
  / t $ ($\,\hbar=c=a=1$) as a function of temperature for a constant $d_{x^2-y^2}$-wave gap
 with phase fluctuations ($\langle n \rangle
  =0.9$). For comparison, we also show the temperature dependence of
  the spin susceptibility $\chi_{spin}$ for a BCS-superconductor with
  a constant gap and for the tight-binding model ($\Delta=0$).}
\label{chi_spin}
\end{figure}

\begin{figure}[t]
\begin{center}
\includegraphics[width=7cm]{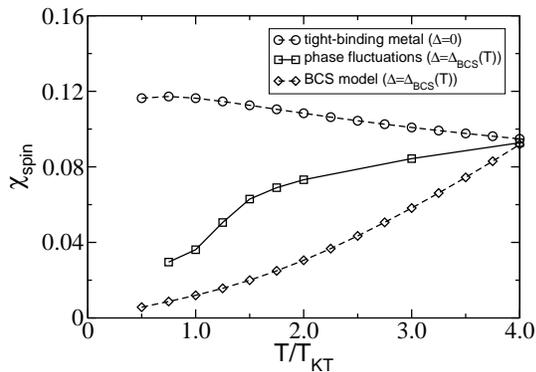}
\end{center}
\caption[]{Spin susceptibility $\chi_{spin}$ in units of $g^2 \mu_B^2
  / t $ ($\,\hbar=c=a=1$) as a function of temperature for a BCS-temperature-dependent $d_{x^2-y^2}$-wave gap
 with phase fluctuations ($\langle n \rangle
  =0.9$). For comparison, we also show the temperature dependence of
  the spin susceptibility $\chi_{spin}$ for a BCS-superconductor and
  for the tight-binding model ($\Delta=0$).}
\label{chi_spin_temp}
\end{figure}

\begin{figure}[t]
\begin{center}
\includegraphics[width=7cm]{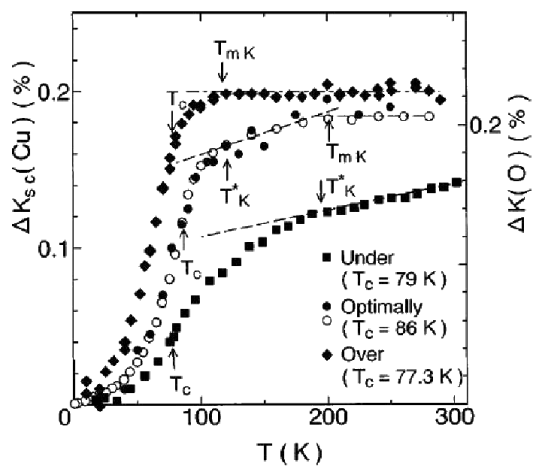}
\end{center}
\caption[]{Experimental results for the c-axis NMR Knight-shift in 
underdoped, optimally doped and overdoped Bi2212 
from Ref.~\onlinecite{is.yo.98}. The temperature variation of the NMR 
Knight-shift scales linearly with the macroscopic magnetic susceptibility 
$\chi_{spin}$. The anomalies indicated by arrows are discussed in the text.}
\label{knight_fig}
\end{figure}

Fig.~\ref{chi_spin} displays the spin susceptibility 
$\chi_{spin}$ as a function of temperature for a constant 
$d_{x^2-y^2}$-wave gap with phase fluctuations at
finite doping ($\langle n \rangle=0.9$). 
In Sec.~\ref{intro}, we have mentioned already that the fluctuation-induced
diamagnetism
is only important below $T \simeq 1.5 \, T_c$
and that, at higher temperatures, $\chi_{Dia}$ is
at least two orders of magnitude smaller than the paramagnetic 
current response of the tight-binding electrons.
Moreover, the experimentally observed diamagnetic
susceptibility close to $T_c$ can be perfectly fitted by an exponential 
function, which also indicates that the fluctuating diamagnetism 
is exponentially suppressed at higher temperatures \cite{se.be.01}.
Furthermore, one obtains that the 
paramagnetic spin response is about 5 times larger than the
paramagnetic current response of the tight-binding electrons.
In addition, the magnetic  current
response is only  weakly temperature dependent.
Hence, we expect that the
temperature dependence of the experimentally observed paramagnetic 
susceptibility is dominated by the spin susceptibility $\chi_{spin}$ 
over almost the the entire pseudogap phase.\footnote{It is known, from
	a variety of experiments (e.g. Fong et. al., PRB {\bf 61}, 14773
		[2000]), that there is an increase of AF correlations in the
	pseudogap if the temperature $T$ is reduced. However, this effect
	has been detected experimentally in the low-doping regime,
	especially for dopings where at low $T$ the AF phase appears. We
	neglect in our
	discussion the direct influence of antiferromagnetic spin
	fluctuations. As discussed in Section I, they may, however, be
	indirectly incorporated in the effective next-neighbor $d_{x^{2} -
		y^{2}}$-pairing interaction used in our BCS-like model (Eqs.
		(1)-(4)).}

In Fig.~\ref{chi_spin} $\chi_{spin}$ is plotted for our
phase-fluctuation model and, additionally, for the
nearest-neighbor tight-binding model and for a BCS-model with
a constant superconducting gap. The phase-fluctuation model
exhibits a characteristic temperature dependence of the spin
susceptibility, which differs qualitatively from the tight-binding model
as well as from the BCS-model. This becomes even more clear 
for a BCS-temperature-dependent gap, as shown in  Fig.~\ref{chi_spin_temp}.
From Fig.~\ref{chi_spin_temp} we can also infer, that a temperature
dependent pairing-gap only slightly modifies the temperature
dependence of $\chi_{spin}$ in our phase-fluctuation model. 
$\chi_{spin}$ always slightly decreases (nearly linear) below $T^*
\equiv T_c^{MF}$ and, then, displays a characteristic downward bending
at $T \simeq 2 \, T_{KT}$.

These results are very similar to the experimentally observed
temperature dependence of the NMR Knight-shift and the
magnetic susceptibility in the pseudogap state of various
underdoped high-$T_c$ compounds
\cite{is.yo.98,zh.cl.99,wa.fu.00,ko.to.01}. For comparison, we
reproduce in Fig.~\ref{knight_fig} the $T$-dependence of the Knight-shift
experimental data from Ref.~\onlinecite{is.yo.98} in underdoped and overdoped Bi2212.
In the pseudogap state, the temperature variation of the NMR Knight-shift
scales linearly with the macroscopic magnetic susceptibility \cite{al.oh.89}.
In underdoped Bi$_2$Sr$_2$CaCu$_2$O$_{8 + \delta}$
(Bi2212) single-crystals, two characteristic temperature scales
can be identified in the temperature dependence of the Knight-shift
\cite{is.yo.98} (see Fig.~\ref{knight_fig}). A higher temperature $T_{mK}$, where the Knight-shift starts 
to decrease from the nearly constant high-temperature value, and a
lower temperature $T^*_{K}>T_c$, where it starts to decrease very steeply \cite{is.yo.98}. 
These two temperatures are, however, difficult to define exactly.
This can be seen from the scaled Knight-shift date of YBCO shown in 
Ref.~\onlinecite{ta.lo.01} and also from our numerical results displayed in 
Fig.~\ref{chi_spin} and Fig.~\ref{chi_spin_temp}, since the change 
as a function of temperature is continuous.

\section{Summary and conclusion}

In the high-$T_{c}$ superconductors, a pseudogap opens at at
temperature $T^{*}$ that is higher than the SC transition
temperature $T_{c}$. There is a rather general consensus that the
pseudogap state is crucial for understanding the microscopic pairing
mechanism, however key issues are unsettled. One of these is whether
it is a precursor state for SC, sharing common features such as pair
formation or whether it is even antagonistic to SC.

At present, there is not yet agreement as to which of these
possibilities is correct. In part, this is because of the difficulty in
determining the experimental consequences of the various theoretical
proposals. This is exactly what we seek to improve here.

We have provided a numerical solution of a simplified model which,
nevertheless, contains key ideas of the phase-fluctuation pseudogap
scenario. Here, the center-of-mass pair-phase fluctuations of a BCS
$d$-wave model were determined from a classical 2D $XY$-action by means
of a Monte-Carlo simulation. Earlier work concerned with the
one-particle excitations have already been found to reproduce salient
features of recent STM studies of Bi2212 and Bi2201
\cite{ec.sc.02} and to explain the change of the quasiparticle
dispersion in crossing $T_{c}$ in underdoped cuprates \cite{ec.ha.03}.
Here, we concentrate on experiments related to two-particle
correlation functions, i.~e.~the optical and magnetic response
functions. They are sensitive, diagnostic tools to search for
normal-state remnants of the infinite d.~c.~conductivity and perfect diamagnetism above $T_{c}$.

In the first part of the present work, we have studied the
frequency-dependent complex conductivity. Our motivation there stems
partly from the work of the Berkeley group\cite{co.ma.99}, which very convincingly
demonstrated that the consequences of partial phase coherence can only
be resolved by high-frequency optical experiments, probing the
short-time dynamics of the phase correlations.

A detailed analysis of the temperature and frequency dependence of the
optical conductivity $\sigma(\omega)=\sigma_1(\omega)+i\sigma_2(\omega)$
revealed a superconducting scaling of $\sigma_2(\omega)$,
which starts already above $T_c$, exactly as observed in
high-frequency microwave conductivity experiments on Bi2212\cite{co.ma.99}. 

Secondly, we calculated the magnetic susceptibility. 
The temperature dependence of 
the uniform static magnetic susceptibility is dominated by the Pauli
spin susceptibility, 
which displayed a very characteristic temperature dependence, independent of 
the details of the gap function used in our model. This temperature dependence is
qualitatively very similar to the experimentally observed change of the Knight-shift 
as a function of temperature in underdoped Bi2212.\\

\section*{Acknowledgments}

We would like to acknowledge useful discussions and comments by D.~J.~Scalapino,
 N.~P.~Ong and E.~Arrigoni. This work was supported
by the DFG Forschergruppe 538 and by the KONWIHR
supercomputing network in Bavaria. 


\end{document}